%% file: kaonfpcp1.tex
\documentclass[12pt]{article}
\usepackage{epsfig,graphicx}
\usepackage{fpcp04}

\input fpcp04-macro
\begin{document}

\Title{Flavor Physics in Kaon Decays}
\bigskip

%
\label{CSKimStart}

%
\author{ Xiao-Gang He\index{He, X.-G.} }

%
\address{NCTS/TPE, Department of Physics\\
National Taiwan University\\
Taipei, Taiwan\\
}

\makeauthor\abstracts{
In this talk I review some topics related to flavor physics in
kaon decays with emphasis on the roles which kaon decays can play in
the determinations of hadronic matrix elements, the
CKM parameters and new physics beyond the Standard Model.
}

\section{Introduction}

Kaon physics has a
glorious history. Parity violation, CP violation, GIM mechanism,
CKM mechanism, Determination of Standard Model (SM) parameters, Hadronic physics
all have imprints of kaon physics. Kaon physics still plays an important
role in particle physics. Some of the main interests today in kaon
physics are: The studies of CP violation and rare decays; Determination of SM
parameters; Low energy tests of QCD, lattice QCD and low energy
effective theories; New physics beyond the SM. There are also many studies of
tests of quantum mechanics
and CPT symmetry in kaon decays. Some of the related work has been reported
at this conference\cite{ktalks}.

Since FPCP04 is a flavor and CP conference, I will mainly discuss
flavor physics in kaon decays with emphases on: $\epsilon_K$ and
$\epsilon'_K/\epsilon_K$ and related hadronic parameters;
Rare kaon decays and the CKM parameters; And rare kaon decays and
new physics beyond the SM.

In the SM, flavor changing neutral current and CP violating
phenomena are all related to the CKM matrix in the charged current
interaction of the W-boson with quarks,
\begin{eqnarray}
L_{int} = -{g\over \sqrt{2}} \bar U \gamma^\mu {1-\gamma_5\over 2}V_{CKM} D W^+_\mu
+ H.C.
\end{eqnarray}
where $V_{CKM}$ is an unitary matrix which are usually written as
\begin{eqnarray}
V_{CKM} = \left ( \begin{array}{ccc} V_{ud}&V_{us}&V_{ub}\\
V_{cd}&V_{cs}&V_{cb}\\
V_{td}&V_{ts}&V_{tb}
\end{array}
\right );
\;\;\begin{array}{l}
V_{us}=s_{12}c_{13}=\lambda,\\
V_{ub}=s_{13}e^{-i\gamma} =A\lambda^3(\rho -i\eta),\\
V_{cb} =
s_{23}c_{13} = A\lambda^2.
\end{array}
\end{eqnarray}

At present, some of the elements are well determined and some of
them are not, such as the parameters $\rho$ and $\eta$. A very
important task for flavor physics is to determine the parameters
$\rho$ and $\eta$. In this talk the following set of numbers
determined from CKMfitter will be taken as reference
values\cite{ckmfitter},
\begin{eqnarray}
&&\lambda = 0.2265^{+0.0025}_{-0.0023},\;\;
A=0.801^{+0.029}_{-0.020},\;\;\rho=0.189^{+0.088}_{-0.070},\nonumber\\
&&\eta
= 0.358^{+0.046}_{-0.042},\;\;
\gamma = 62^{+10}_{-12}(\mbox{degree}).
\end{eqnarray}

\section {Kaon Decays and the Standard Model}

\noindent {\bf 2.1 The CP violating parameters $\epsilon_K$ and
$\epsilon'_K/\epsilon_K$}

$\epsilon_K$ and $\epsilon'_K/\epsilon_K$ are the two famous
quantities parameterize CP violation in $K^0-\bar K^0$ mixing and
neutral kaon to two pion decays. These and a closely associated
parameter $\Delta m_K$  play important roles in understanding CP
violation and flavor changing interactions. These parameters are,
now, measured to good precisions with\cite{pdg} $\epsilon_K =
(2.284\pm 0.014)\times 10^{-3}e^{i43.51^o\pm 0.05^o}$,
$|\epsilon'_K/\epsilon_K| = (1.67\pm 0.23)\times 10^{-3}$, and
$\Delta m_K = (3.483\pm 0.006)\times 10^{-12}$ MeV.

The short distance contributions, at the quark level, to
$\epsilon_K$ and $\epsilon'_K/\epsilon_K$ have been calculated in
the SM and are well understood. But there are large uncertainties
in hadronic matrix elements such as $\hat B_K$, $R_6$ and $R_8$
which cause uncertainties in the determination of the CKM
parameters using these observables.

The parameters $\epsilon_K$ and $\Delta m_K$ are related to the
$K^0 -\bar K^0$ mixing parameters $M_{12}$ with $\epsilon_K = {Im
(M_{12})e^{i\pi/4} /\sqrt{2}\Delta m_K}$ and $\Delta m_K = 2
Re(M_{12})$. The short distance contribution comes from the
``box'' diagram\cite{box} which is given by
\begin{eqnarray}
M_{12} = {G_F^2 m^2_W m_K\over 12 \pi^2} f^2_K \hat B_K \left [
\eta_{cc} S(x_c, x_c) (\lambda_{c})^2 + \eta_{tt} S(x_t,x_t)
(\lambda_t)^2 +2\eta_{ct} S(x_c, x_t) \lambda_c \lambda_t\right ].
\end{eqnarray}
The function $S(x,y)$ and the QCD correction factors $\eta_{ij}$
are all known\cite{buras1}. The parameter $\hat B_K$ is defined as

\begin{eqnarray}
<\bar K^0 | \bar s \gamma_\mu (1-\gamma_5) d \bar s \gamma^\mu
(1-\gamma_5) d| K^0> = {8\over 3} f^2_K m^2_K \hat B_K.\nonumber
\end{eqnarray}
In the vacuum saturation and factorization approximation, $\hat B_K =1$.
Lattice calculation gives\cite{lattice1} $0.86\pm 0.06\pm 0.14$.

\begin{figure}[!htb]
\begin{center}
\caption{Constraints\cite{ckmfitter} on CKM parameters. }
\vspace{0.3cm}
\includegraphics[width=6cm]{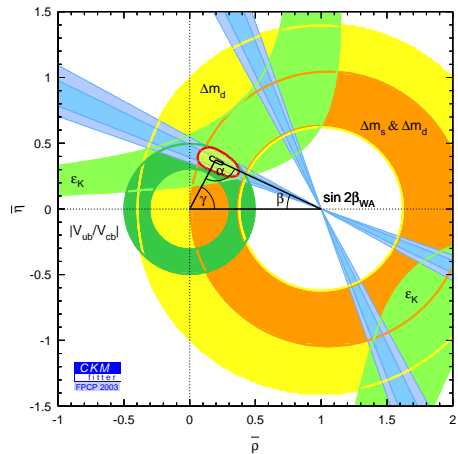}%
\end{center}
\label{constraints}
\end{figure}

Till 1999, before the measurement of $\epsilon_K'/\epsilon_K$,
$\epsilon_K$ was the only measured CP violating quantity. It
played a crucial role in determining the CKM parameter $\eta$ as
can be seen in Fig. 1. There are now more hadronic matrix element
independent determinations of CKM parameters such as
$|V_{ub}/V_{cb}|$ and $\sin(2\beta)$ from $B$ decays. Using CKM
parameters determined from other processes, one can study the
hadronic parameter $\hat B_K$ in a more precise way. Requiring
that the observed $\epsilon_K$ be the same as data, one can
determine the parameter $\hat B_K$.  The UTfit group has done such
an analysis and obtained\cite{utfit}
\begin{eqnarray}
\hat B_K = 0.65 \pm 0.10.
\end{eqnarray}
This is in agreement with lattice calculations within error bars.
If there are new physics beyond the SM, the fit may change. But
the parameters can be constrained.

The quantity $\Delta m_K$ is less well understood. If one uses the
central values of the CKM parameters, using eq.(4) one would
obtain $\Delta m_K|_{SD} = 2.72 \hat B_K \times 10^{-12}$ MeV
which is substantially smaller than the experimental value.
Theoretical calculation of $\Delta m_K$ is indeed more difficult
since there are long distance contributions. For example, exchange
of $\pi$, $\eta$ and $\eta'$\cite{don}, can make a contribution
given in the following\cite{h1}
\begin{eqnarray}
&&2m_K \, {\rm Re} (M_{12}^{one})= {|\langle \pi^0|H_W|K^0\rangle|^2\over m_K^2-m_\pi^2}
\left[ 1+ {m_K^2-m_\pi^2\over m_K^2-m_\eta^2} \left( {1+\delta+\delta^{gg}
\over \sqrt{3}} \mbox{cos}\theta
+{2\sqrt{2}\over \sqrt{3}}(\rho+r^{gg}) \mbox{sin}\theta\right)^2 \right .\nonumber\\
&&+ \left .
{m_K^2-m_\pi^2\over m_K^2-m_{\eta'}^2} \left( {1+\delta+\delta^{gg}\over \sqrt{3}} \mbox{sin}\theta - {2\sqrt{2}
\over \sqrt{3}} (\rho+r^{gg}) \mbox{cos}\theta \right)^2 \right].\nonumber
\end{eqnarray}
where $\delta = 0.17$, $\rho \approx 1$ which parameteriz
the SU(3) and nonet breaking parameters, and $\delta^{gg} = 0.96a_2$,
$r^{gg} =-0.48 a_2$ come from $s\to d gg$ contribution.
Fitting $K_L \to \gamma\gamma$ data, $a_2 = 0.17$, $\Delta
m^{one}_K = -0.9\times 10^{-12}$ MeV which make the total $\Delta m_K$
even smaller.
$\Delta m_K$ is not well understood. More study is needed.

There is also uncertainties in $\epsilon'_K/\epsilon_K$ due to our
limited understanding of hadronic matrix elements. In the SM, there are
tree, penguin and box contributions to $\epsilon'_K/\epsilon_K$.
One obtins\cite{epsilonp}
\begin{eqnarray}
&&{\epsilon'_K \over \epsilon_K} = Im\lambda_t \left ( 18.7 R_6
(1-\Omega_{IB}) - 6.9 R_8 - 1.8\right ) \left [
{\Lambda^4_{\overline{SM}}\over 340\mbox{MeV}}\right ]\nonumber
\end{eqnarray}
$\Omega_{IB} = 0.06\pm 0.08$ comes from isospin
breaking\cite{don1}. The definition of $R_{6,8}$ are given by,
\begin{eqnarray}
&&<\pi\pi|Q_6|K>_0= -4\sqrt{3\over 2}(f_K-f_\pi) \left
({m^2_K\over
121\mbox{MeV}}\right )^2R_6\mbox{GeV}^3,\nonumber\\
&&<\pi\pi|Q_8|K>_2 =\sqrt{3}f_\pi\left ({m^2_K\over
121\mbox{MeV}}\right )^2 R_8\mbox{GeV}^3.\nonumber
\end{eqnarray}
Lattice calculation gives\cite{lattice2} $R_8 = 1.00\pm 0.20$.
$R_6$ is much harder to calculate. There are also other calculations.
For example large N approximation gives $R_6 = R_8$.

Again, requiring the data to be reproduced and using CKM matrix elements
determined from other processes, one can obtain more information about the
hadronic matrix elements. We show the correlation of $R_6$ and $R_8$ in
Fig. 2.
Any model calculations for these hadronic parameters can therefore
be tested.

\begin{figure}[!htb]
\begin{center}
\caption{Correlation\cite{ckmfitter,bb6} between $R_6$ and $R_8$.
} \vspace{0.3cm}
\includegraphics[width=6cm]{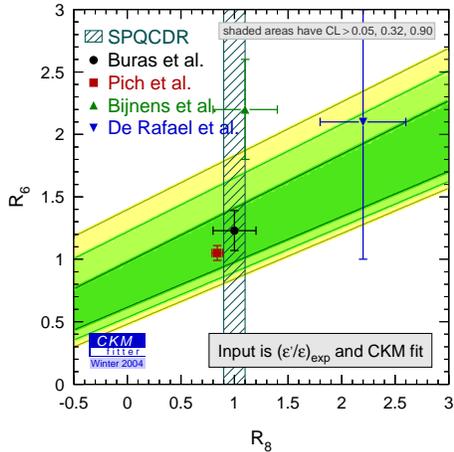}%
\end{center}
\label{correlation}
\end{figure}

We would like to emphases that present knowledge of CKM parameters
determined from other processes can provide important information
about hadronic physics at low energies. There are still rooms for
new physics. One can also take
the same attitude to study hadronic parameters in B decays.
\\

\noindent {\bf 2.2 Rare Kaon Decays In The Standard Model}

I now discuss the rare kaon decays: $K_L\to \mu^+\mu^-,\; K_L\to
\pi^0 e^+e^-,\;K\to \pi \nu \bar \nu$ in the SM. The experimental
measurements are summarize in Table 1.

\begin{table}[htbp]
\begin{center}
\begin{tabular}{|l|l|l|l|l|}
\hline
 & $K_L\to \mu^+\mu^-$ & $K_L \pi^0 e^+ e^-$ & $K^+\to \pi^+ \nu\bar \nu$ &
$K_L \to \pi^0 \nu \bar \nu$ \\ \hline
Br&$(7.27\pm 0.14)\times 10^{-9}$ & $< 2.8 \times 10^{-10}$ &
$(1.47^{+1.30}_{-0.89})\times 10^{-10}$ & $< 5.6\times 10^{-7}$\\
\hline
& $K^+ \to \pi^+ \mu^+ e^-$
& $K^+\to \pi^+ \mu^- e^+$
& $K_L \to \pi^0 \mu^\pm e^\mp$
& $K_L \to \mu^\pm e^\mp$\\
\hline
Br&
$< 1.2\times 10^{-11}$& $< 5.2\times 10^{-10}$
& $< 3.4\times 10^{-10}$&
$< 4.7\times 10^{-12}$\\
\hline

\end{tabular}
\caption{Experimental data on branching ratios of rare kaon
decays\cite{ktalks,pdg}.}
\label{rare}
\end{center}
\end{table}

{\bf $K_L \to \mu^+ \mu^-$}. There are long and short distance
contributions to this rare decay. The long distance contribution
is mainly from two photon intermediate state, $K_L \to \gamma
\gamma \to \mu^+ \mu^-$. The absorptive contribution, with the two
photons on-shell, is calculated to be $(7.07\pm 0.18)\times
10^{-9}$ almost saturate the total branching ratio\cite{valencia}.
The dispersive contribution, with the two photons off-shell, is
not well understood. The short distance and dispersive
contributions to the branching ratio together should be less than
$2.5\times 10^{-9}$. The short distance contribution from box and
Z penguin diagrams in the SM is given by $B(K_L\to \mu^+
\mu^-)_{SD} = 1.6\times 10^{-9}A^4(\rho_o' - \rho)^2 = (0.8\pm
0.3)\times 10^{-9}$. Here we have used the SM calculation of
$\rho_0' = 1.2$. We emphases that since the dispersive
contribution is not well known, one should not use this process as
a good one to determine CKM parameters. But it can provide
non-trivial constraints on new physics beyond the SM.

{\bf $K_L \to \pi^0 e^+ e^-$}. There are CP conserving and CP
violating contributions to this proccess\cite{don2,isidori}. The
main CP conserving contribution is from two photon intermediate
state of the type, $K_L \to \pi^0 \gamma\gamma \to \pi^0 e^+ e^-$.
This contribution alone has been shown to give a small branching
ratio ($< 3\times 10^{-12}$). There are two types of CP violating
contributions, one from $K_L$ and $K_S$ mixing with $K_S$ as an
intermediate state of the type, $K_L \to K_S \to \pi^0 e^+e^-$,
and another from short distance contribution. The short distance
contribution alone is predicted to be small (about $4.7 \times
10^{-12} $). Using the recently measured\cite{ktalks,na48}
$B(K_S\to \pi^0 e^+e^-)_{m_{ee} > 165 \mbox{MeV}}
=(3.0^{+1.5}_{-1.2}\pm 0.2)\times 10^{-9}$. and $B(K_S\to \pi^0
\mu^+\mu^-) = (2.9^{+1.4}_{-1.2}\pm 0.2)\times 10^{-9}$, and
assuming vector dominance\cite{isidori}, $B(K_S \to \pi^0 e^+
e^-)$ is shown to be $4.6 a^2_S \times 10^{-9}$. $a_s$ to be
positive is favored and $|a_S|$ is determined to be $1.2 \pm 0.2$.
One obtains\cite{isidori},
\begin{eqnarray}
B(K_L\to \pi^0 e^+e^-)_{SM} = B_{CPC} + (15.3 a_S^2 - 35 A^2 \eta
a_S + 74.4 A^4 \eta^2) \times 10^{-12}.
\end{eqnarray}
This results in
$B(K_L\to \pi^0 e^+e^-)_{SM} = (3.7\pm 0.4)\times 10^{-11}$,
and
$B(K_L\to \pi^0 \mu^+\mu^-)_{SM} = (1.5\pm 0.3)\times 10^{-11}$.
These values are much smaller than the current upper bounds. Since model
calculations are involved, the above should be taken to be some reasonable
estimate. The process $K_L \to \pi^0 e^+ e^-$ can be used to constrain
possible new physics, but should not to be considered
as a good process for CKM parameter determination.

{\bf $K\to \pi \nu \bar \nu$}. These are considered golden modes
for the determination of the CKM parameters since hadronic
uncertainties are eliminated\cite{buras2}. In the SM, $K^+\to
\pi^+ \nu\bar \nu$ is given by
\begin{eqnarray}
&&B(K^+\to \pi^+\nu\bar\nu)  = {r_{K^+} \alpha^2\over 2 \pi^2
s^4_W |V_{us}|^2} \sum_{e,\mu\tau}|\Delta^l_K|^2B(K^+\to \pi^0 e^+
\nu_e),\;\;\Delta_K^l = \lambda_c X^l_{NL} + \lambda^X
\eta_X X_0(x_t),\nonumber
\end{eqnarray}
$X_0(x)$, $X_{NL}^l$ and $\eta_X$ are known\cite{buras3}, and
$r_{K^+} = 0.901$\cite{marciano}.
The uncertainty for the branching ratio is
at a few percent level.

The branching ratio is some times written as\cite{buras2}
\begin{eqnarray}
&&B(K^+ \to \pi^+ \nu \bar \nu) = 8.9\times 10^{-10}
A^4 {1\over 3} [(\eta^2 + (\rho_0^e - \rho)^2) + (\eta^2 +
(\rho_0^\mu -
\rho)^2) +(\eta^2 + (\rho^\tau_o - \rho)^2)]\nonumber\\
&& = 8.9\times 10^{-10} A^4 [\eta^2 + (\rho_0 -\rho)^2 + {2\over
9} (\rho^e_0 - \rho^\tau_0)^2],\nonumber
\end{eqnarray}
where $\rho_0 = (\rho^e_0 + \rho^\mu_0 + \rho^\tau_0)/3 \approx 1.4$.
The term  $(2/9)(\rho^e_0 - \rho^\tau_0)^2$ is small ($\sim 3\%$)
and can be neglected. In the SM the branching ratio is predicted
to be $(7.8\pm 1.2)\times 10^{-11}$. There are experimental
measurements for this process combining E787\cite{e787} and E949
data\cite{e949}, the branching ratio is\cite{e949}
$(1.47^{+1.3}_{-0.89})\times 10^{-10}$. The central value is about
two times larger than the SM prediction. Of course the error bar
is too large to draw definitive conclusion about whether new
physics is needed. But future improved experiments will provide us
with more information.

The branching ratio of $K_L\to \pi^0\nu\bar\nu$ in the SM is given by
\begin{eqnarray}
&&B(K_L \to \pi^0\nu\bar\nu) = {r_{K_L} \alpha^2\over 2 \pi^2 s^4_W
|V_{us}|^2}{\tau_{K_L}\over \tau_{K^+}}
\sum_{e,\mu\tau}|Im(\Delta^l_K)|^2B(K^+\to \pi^0 e^+ \nu_e),\nonumber\\
&&\nonumber\\
 &&= 3.34\times 10^{-5}(\eta_X X_0(x_t))^2 |V_{cb}|^4
\eta^2.\nonumber
\end{eqnarray}
$r_{K_L} = 0.994$. $B(K_L\to \pi^0 \nu \bar \nu)$ is
predicted to be $(3.0\pm 0.6)\times
10^{-11}$.

\begin{figure}[!htb]
\begin{center}
\caption{$K\to \pi \nu\bar \nu$ for CKM triangle.}
\vspace{0.3cm}
\includegraphics[width=6cm]{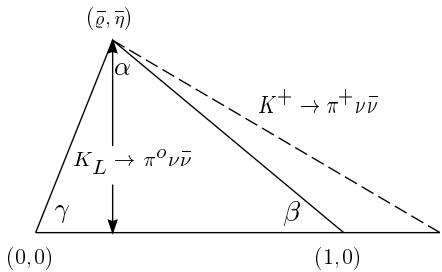}%
\end{center}
\label{knunu}
\end{figure}

It is clear that the processes $K^+\to \pi^+ \nu\bar \nu$ and
$K_L\to \pi^0 \nu \bar \nu$ can play a very important role in the
determination of the unitarity triangle which is schematically
shown in Fig. \ref{knunu}. The phase $\gamma$ can be expressed
as\cite{buras2}
\begin{eqnarray}
\cot\gamma =\sqrt{1\over B_2} (A^2 \eta_X X_0(x_t) -
\sqrt{B_1-B_2} + P_c(X)),
\end{eqnarray}
making a direct measurement of $\gamma$.
Here $B_1 =
{B(K^+\to \pi^+ \nu\bar \nu)/\kappa_+}$, $B_2 =
{B(K_L\to \pi^0 \nu\bar \nu)/ \kappa_L}$
with $\kappa_+ = r_{K^+} (3\alpha^2 B(K^+\to \pi^+ \nu\bar\nu))/(2
\pi^2 s^4_W)\lambda^8$, $\kappa_L = \kappa_+ (r_{K_L}/
r_{K^+}) (\tau(K_L)/\tau(K^+))$, and
$P_c(X) = (1/3) (2X_{NL}^e +X_{NL}^\tau)$.

\section{Kaon Decays and New Physics}

Rare kaon decays being rare are very sensitive to new physics
beyond the Standard Model\cite{beyondsm}. There are two aspects in
discussing new physics. On the one hand, one can try to explain
processes where deviations from the SM already shown up. And on
the other hand, one can use the current measurements on rare
processes to constrain model parameters. As have been mentioned
earlier that indeed the current experimental central value for the
branching ratio of the rare kaon decay $K^+\to \pi^+ \nu \bar \nu$
is almost two times the SM prediction. This needs to be explained.
It may be an indication of new physics. Of course one needs to be
aware that the error bar is still large, a definitive conclusion
cannot be drawn. It is, however, interesting to see if new physics
is able to produce the central value.

When going beyond the SM, there are new interactions which can
induce rare kaon decays and therefore modify the SM predictions\cite{beyondsm}.
As illustrations, I list some of the model predictions for
$R(K_i \to \pi_i \nu\bar \nu) =
B(K_i\to \pi_i \nu\bar \nu)/B(K_i\to \pi_i \nu\bar \nu)$ taking into account of
constraints from other kaon decay processes discussed previously.

\begin{itemize}

\item SUSY with minimal flavor violation. This is a class of
models with minimal SUSY extension of the SM assuming that there
are no new flavor and CP violating sources. The only source of
flavor and CP violation comes from the CKM sector. Even with such
limited extension there are new contributions to $K\to \pi \nu
\bar \nu$ coming from chargino and charged Higgs exchanges. These
new contributions are, however, stringently constrained. $R(K\to
\pi \nu \bar \nu)$ are constrained to be\cite{mfv}: $0.65<R(K^+\to
\pi^+ \nu \bar \nu) < 1.02$, $0.41 < R(K_L\to \pi^0 \nu \bar \nu)
< 1.03$.

\item General MSSM. In more general supersymmetric minimal SM,
there are more flavor and CP violating sources. Sizeable
enhancements to $K\to \pi \nu \bar \nu$ can come from chargino
exchange with large up-squark mixing. $R(K^+\to \pi^+ \nu\bar
\nu)$ and $R(K_L\to \nu \bar \nu)$ can reach $2.1$ and $ 4$,
respectively\cite{gsusy}.

\item R-parity violating model. In this case there are tree level
contributions to $K\to \pi \nu \bar \nu$ by exchange sqaurks. The
$K^+\to \pi^+ \nu\bar \nu$ can easily produce the central value of
current data, and the value for $R(K_L\to \pi^0 \nu \bar \nu)$ can
be as large as 70\cite{rparity,rparity1}.

\item Modified Left-Right model. In a recent model proposed to
solve the $A^b_{FB}$ problem, the third generation of right-handed
fermions are assumed to have different interactions compared with the
first two generatios. As a result, exchange of a
heavy $Z'$ boson can induce tree level flavor changing neutral
current. There are also sizeable loop effects on $K\to \pi \nu
\bar \nu$. After constraints from other processes are taken into
account, one finds that\cite{hv} $R(K^+\to \pi^+ \nu \bar \nu)$
can be as large as 2, and $R(K_L\to \pi^0 \nu \bar \nu)$ can be as
large as $5$.

\item Little Higgs model. Little Higgs models provides a new
solution to hierarchy problem in the SM. One of the feature of
this type of models is the existence of a vector like top quarks.
Exchange of such particle at the loop level, large branching ratio
for $K_L\to \pi^0 \nu \bar \nu$, as large as 3 times of the SM
prediction\cite{little}, can be generated.

\item Isosinglet down quark model. In this type of models, it is
possible to induce flavor changing neutral current by exchange $Z$
at tree level. $R(K^+\to \pi^+\nu\bar \nu)$ and $R(K_L\to
\pi^0\nu\bar\nu)$ can reach $2$ and $3$,
respectively\cite{rparity,isosinglet}.
\end{itemize}

$K\to l\bar l'$, $K \to \pi l \bar l',\; \pi \nu \bar \nu'$ with
$l\neq l'$. These processes
are forbidden in the SM. Observations of these modes are clear
indications of new physics beyond the SM. Non-observations of
these decays can put stringent constraints on new physics beyond
the SM.

In order for these decays to occur, there must be flavor changing
neutral current in the lepton sector. It can be achieved in many
ways. For example, in Left-Right symmetric models, through
exchanges of heavy neutrinos, sizeable $K\to l \bar l'$ and $K\to
\pi l \bar l'$ can be generated\cite{lrmodel}. In R-parity
violating models, there are even tree level effects which can
induce $K\to l \bar l' (\nu \bar \nu'),\;\; \pi l \bar l'(\pi \nu
\bar \nu')$\cite{rparity,rparity1}. For illustration, I discuss in
more detail for $K\to \pi \nu_i \bar \nu_j$ in R-parity violating
model

There are several types of R-parity violating interactions. The
one most directly related to the discussion here is
the $R$-parity violating interaction terms : ${\cal W} =
\lambda^\prime_{ijk}\hat L_i \hat Q_j \hat D^c_k$.
For $K_L \to \pi^0 \nu_i \bar \nu_j$ decay, we have
\begin{eqnarray}
{{\rm Br} (K_L\to \pi^0 \nu \bar \nu)\over {\rm Br}(K^+ \to \pi^0
e^+\nu)} &=& \kappa_L \left [\sum_{\ell=e,\mu,\tau}\mid {\rm
Im}(\Delta^{SM}_{K} + \Delta^{R}_{K_L \ell\ell})\mid^2
 +\sum_{i\neq i'}
\mid \Delta^{R}_{K_L ii'} \mid^2 \right ]\nonumber
\end{eqnarray}
\begin{eqnarray}
\Delta^{R}_{K_Lii'} = {\pi s^2_W\over \sqrt{2}G_F\alpha}\left [
{\lambda'_{i'j1}\lambda^{'*}_{ij2}\over 2m^2_{\tilde d_L^j}} -
{\lambda'_{i'2k}\lambda^{'*}_{i1k}\over 2 m^2_{\tilde d^k_R}}
-{\lambda'_{i'j2}\lambda^{'*}_{ij1}\over 2m^2_{\tilde d_L^j}} +
{\lambda'_{i'1k}\lambda^{'*}_{i2k}\over 2 m^2_{\tilde d^k_R}}
\right ]. \label{klpinn}\nonumber
\end{eqnarray}
It is interesting to note that for $i=i'$, the decay is $CP$
violating. But for $i\neq i'$, it is not necessarily $CP$
violating since couplings can all be real to obtain non-zero decay
rates. This is very different from SM\cite{gross}.

We list the constraints on various couplings in Table
2.
We see that stringent constraints can be obtained for
the couplings.

\begin{table}
\begin{center}
\begin{tabular}{|c|c|c|}
\hline
Couplings & bounds    & source \\
\hline $ \mid \lambda'_{11j}\lambda'_{32j}\mid $ & $ 0.89 \times
10^{-5} $
& $ K^+ \to \pi^+ \nu \bar \nu' $ \\
\hline $ \mid \lambda'_{21j}\lambda'_{32j}\mid $ & $ 0.89 \times
10^{-5} $
& $ K^+ \to \pi^+ \nu \bar \nu' $ \\
\hline $ \mid \lambda'_{31j}\lambda'_{12j}\mid $ & $ 0.89 \times
10^{-5} $
& $ K^+ \to \pi^+ \nu \bar \nu' $ \\
\hline $ \mid \lambda'_{31j}\lambda'_{22j}\mid $ & $ 0.89 \times
10^{-5} $
& $ K^+ \to \pi^+ \nu \bar \nu' $ \\
\hline $\mid \lambda'_{i^\prime j 2}\lambda'_{ij1}\mid $
& $
0.89 \times 10^{-5} $
& $ K^+ \to \pi^+ \nu \bar \nu' $ \\
\hline $\mid \lambda'_{i^\prime j 2}\lambda'_{ij1}\mid $
 & $
0.89 \times 10^{-5} $
& $ K^+ \to \pi^+ \nu \bar \nu' $ \\
\hline $ \mid \lambda'_{1j1}\lambda'_{2j2}\mid $ & $ 1.2 \times 10^{-6} $
& $ K^+ \to \pi^+ \mu^+ e^- $ \\
\hline $ \mid \lambda'_{1j1}\lambda'_{2j2}\mid $ & $ 0.8 \times 10^{-6} $
& $ K_L \to \mu^+ e^- $ \\
 \hline
\end{tabular}
\caption{ Upper bounds on the R-parity violating couplings with
$m_{\tilde f_{j}} = 100$ GeV.}
\end{center}
\label{rpv_bound}
\end{table}

\section{ Summary}

Kaon physics has played and is playing an important role in our
understanding of fundamental interactions. There are many different
aspects of the Standard Model and models beyond the SM can be
studied using kaon decays.

Many of the kaon decay processes involve hadronic parameters which are
difficult to study theoretically due to non-perturbative QCD effects.
However, at present many of the electroweak parameters, such as
CKM parameters
are well determined from other processes, hadronic parameters
can now be studied in detail experimentally in kaon decays, such
as $B_K$, $R_{6,8}$ and $\Delta m_K$. Such study can provide good tests for
lattice and other low energy effective theories.

There are also some kaon decays which are free from uncertainties in
hadronic physics, such as
$K^+\to \pi^+ \nu\bar \nu$ and $K_L\to \pi^0 \nu \bar \nu$. These processes
can therefore provide clean determinations for the fundamental CKM
parameters in the SM. Once the branching ratios of these processes
are measured, the Standard Model can be tested.

At present there is no inconsistence with Standard Model in kaon
decays, although the central value for the branching ratio of
$K^+\to \pi^+\nu\bar \nu$ is about two times of the SM prediction.
But the error bar is too large to draw firm conclusion. Should the
current central value will be confirmed, new physics is required
to explain the difference.

When going beyond the SM, there are new interactions which can
induce processes which are forbidden in the SM.
We have seen that rare kaon decays, such as $K_L \to \mu e,
\pi l \bar l', \pi \nu\bar\nu'$  are sensitive to
new physics and can provide stringent constraints on new physics
parameters. Current experimental bounds still allow
rooms for new physics which modify the
SM predictions substantially. Future measurements can provide us with
more information.

We have heard from other talks\cite{ktalks} that in the near
future several experiments will be in operation, such as  NA48 in
CERN, E391 in KEK, Japan JPARC in Japan, KOPIO in BNL and etc. It
can be expected that significant progresses will be made in the
area of kaon physics. Kaon physics will continue to have a prominent status in
the study of flavor and CP violation.

\section*{Acknowledgments}
I thank N. G. Deshpande, D. K. Ghosh, C.-S. Huang, X.-Q. Li, G. Valencia
and Y.-L. Wang for collaborations for some of the work report here.
I also thank the conference organizer for the invitation.

%
\label{CSKimEnd}

\end{document}

%% file: fpcp04-macro.tex
\def\beq{\begin{equation}}
\def\eeq#1{\label{#1}\end{equation}}
\def\eeqn{\end{equation}}

\def\beqa{\begin{eqnarray}}
\def\eeqa#1{\label{#1}\end{eqnarray}}
\def\eeqan{\end{eqnarray}}

\let\bar=\overbar

\def\Dslash{\not{\hbox{\kern-4pt $D$}}}
\def\dslash{\not{\hbox{\kern-2pt $\del$}}}

\def\msb{{\bar{\ssstyle M \kern -1pt S}}}

\def\BB0bar{B^0 {\overline B}^0}
\def\BB0dbar{B_d^0 {\overline B}_d^0}
\def\BB0sbar{B_s^0 {\overline B}_s^0}

\RequirePackage{xspace}

\usepackage{relsize}
\def\babar{\mbox{\slshape B\kern-0.1em{\smaller A}\kern-0.1em
    B\kern-0.1em{\smaller A\kern-0.2em R}}}

\def\Kbar  {\kern 0.2em\overline{\kern -0.2em K}{}\xspace}

\def\Kz    {\ensuremath{K^0}\xspace}
\def\Kzb   {\ensuremath{\Kbar^0}\xspace}
\def\KzKzb {\ensuremath{\Kz \kern -0.16em \Kzb}\xspace}
\def\Kp    {\ensuremath{K^+}\xspace}
\def\Km    {\ensuremath{K^-}\xspace}

\def\KpKm  {\ensuremath{\Kp \kern -0.16em \Km}\xspace}

\def\Dbar    {\kern 0.2em\overline{\kern -0.2em D}{}\xspace}

\def\Dz      {\ensuremath{D^0}\xspace}
\def\Dzb     {\ensuremath{\Dbar^0}\xspace}
\def\DzDzb   {\ensuremath{\Dz {\kern -0.16em \Dzb}}\xspace}
\def\Dp      {\ensuremath{D^+}\xspace}
\def\Dm      {\ensuremath{D^-}\xspace}

\def\DpDm    {\ensuremath{\Dp {\kern -0.16em \Dm}}\xspace}

\def\Bbar    {\kern 0.18em\overline{\kern -0.18em B}{}\xspace}

\def\BB      {\ensuremath{B\Bbar}\xspace} 
\def\Bz      {\ensuremath{B^0}\xspace}
\def\Bzb     {\ensuremath{\Bbar^0}\xspace}
\def\BzBzb   {\ensuremath{\Bz {\kern -0.16em \Bzb}}\xspace}
\def\Bu      {\ensuremath{B^+}\xspace}
\def\Bub     {\ensuremath{B^-}\xspace}

\def\BpBm    {\ensuremath{\Bu {\kern -0.16em \Bub}}\xspace}

\mathchardef\Upsilon="7107
\def\Y#1S{\ensuremath{\Upsilon{(#1S)}}\xspace}

\mathchardef\Deltares="7101
\mathchardef\Xi="7104
\mathchardef\Lambda="7103
\mathchardef\Sigma="7106
\mathchardef\Omega="710A

\def\Deltabar{\kern 0.25em\overline{\kern -0.25em \Deltares}{}\xspace}
\def\Lbar{\kern 0.2em\overline{\kern -0.2em\Lambda\kern 0.05em}\kern-0.05em{}\xspace}
\def\Sigbar{\kern 0.2em\overline{\kern -0.2em \Sigma}{}\xspace}
\def\Xibar{\kern 0.2em\overline{\kern -0.2em \Xi}{}\xspace}
\def\Obar{\kern 0.2em\overline{\kern -0.2em \Omega}{}\xspace}
\def\Nbar{\kern 0.2em\overline{\kern -0.2em N}{}\xspace}
\def\Xb{\kern 0.2em\overline{\kern -0.2em X}{}\xspace}

\newcommand{\tev}{\ensuremath{\mathrm{\,Te\kern -0.1em V}}\xspace}
\newcommand{\gev}{\ensuremath{\mathrm{\,Ge\kern -0.1em V}}\xspace}
\newcommand{\mev}{\ensuremath{\mathrm{\,Me\kern -0.1em V}}\xspace}
\newcommand{\kev}{\ensuremath{\mathrm{\,ke\kern -0.1em V}}\xspace}
\newcommand{\ev}{\ensuremath{\mathrm{\,e\kern -0.1em V}}\xspace}
\newcommand{\gevc}{\ensuremath{{\mathrm{\,Ge\kern -0.1em V\!/}c}}\xspace}
\newcommand{\mevc}{\ensuremath{{\mathrm{\,Me\kern -0.1em V\!/}c}}\xspace}
\newcommand{\gevcc}{\ensuremath{{\mathrm{\,Ge\kern -0.1em V\!/}c^2}}\xspace}
\newcommand{\mevcc}{\ensuremath{{\mathrm{\,Me\kern -0.1em V\!/}c^2}}\xspace}

\def\mus  {\ensuremath{\rm \,\mus}\xspace}

\def\mus        {\ensuremath{\,\mu{\rm s}}\xspace}    

\def\to                 {\ensuremath{\rightarrow}\xspace}

\def\pep2{PEP-II}

\def\gsim{{~\raise.15em\hbox{$>$}\kern-.85em
          \lower.35em\hbox{$\sim$}~}\xspace}
\def\lsim{{~\raise.15em\hbox{$<$}\kern-.85em
          \lower.35em\hbox{$\sim$}~}\xspace}

\xspace

\def\jetset74   {\mbox{\tt Jetset \hspace{-0.5em}7.\hspace{-0.2em}4}\xspace}


%% file: kaonfpcp1.bbl
\begin{thebibliography}{99}


\bibitem{ktalks}  S. Glazov, to be published in FPCP04 proceedings;
T. Komatsubara, to be published in FPCP04 proceedings;
F. Bossi, to be published in FPCP04 proceedings.

\bibitem{ckmfitter} J. Charles et al., hep-ph/0406184.

\bibitem{bb6} D. Becirevic (for SPQCDR Collaboration), Nucl. Phys.
Proc. Suppl. {\bf 119}, 359(2003); A. Buras and M. Jamin, JHEP
{\bf 0401}, 048(2004); E. Pallante, A. Pich and I. Scimemi, Nucl.
Phys. {\bf 617}, 441(2001); J. Bijnens, E. Gamiz and J. Prades,
JHEP {\bf 0110}, 009(2001); T. Hambye, S. Peris and E. de Rafael,
JHEP {\bf 0305}, 027(2003).

\bibitem{pdg} S. Eidelman et al., Particle Data Group, Phys. Lett. {\bf
B592}, 1(2004).

\bibitem{box} T. Inami and C.-S. Lim, Prog. Theor. Phys. {\bf 65},
297(1981); Erratum-ibid, {\bf 65}, 1772(1981).

\bibitem{buras1} S. Herrlich and U. Nierste, Nucl. Phys. {\bf B419}, 292(1994);
Phys. rev. {\bf D52}, 6505(1995); Nucl. Phys. {\bf B476}, 27(1996);
A. Buras, M. Jamin and P. Weisz, Nucl. Phys. {\bf B347}, 491(1990);

\bibitem{lattice1} V. Lubicz, in proceedings of Lattice04, Fermilab, June 21-26,2004).

\bibitem{utfit} M. Ciuchini et al., hep-ph/0307195.

\bibitem{don} J. Donoghue, B. Holstein and Y.-C. Lin, Nucl. Phys. {\bf B277},
651(1986).

\bibitem{h1} X.-G. He, C.-S. Huang and X.-Q. Li, Phys.
Rev. {\bf D67}, 096005(2003).

\bibitem{epsilonp}
S. Bosch et al., Nucl. Phys. {\bf B565}, 3(2000);
A. Buras et al., Nucl. Phys. {\bf B 592}, 55(2001);
A.Buras and M. Jamin, JHEP {\bf 0401}, 048(2004).

\bibitem{don1}
J. Donoghue, E. Golowich, B. Holstein and J. trampetic, Phys. Lett.
{\bf B179}, 361(1986), Erratum-ibid. {\bf B188}, 511(1987);
A. Buras and J. Gerard, Phys. Lett. {\bf B192}, 156(1987);
E. Pallante, A. Pich and I. Scimemi, Nucl. Phys. {\bf B617}, 441(2001).

\bibitem{lattice2} D. Becirevic (for the SPQCDR), Nucl. Phys. Proc. Suppl.
{\bf 119}, 359(2003) [arXiv: hep-let/02091360.


\bibitem{valencia} L. Littenberg and G. Valencia, Ann. Rev. Nucl. Part. Sci.
{\bf 43}, 729(1993).

\bibitem{don2} J. Donoghue, B. Holstein and G. Valencia,
Phys. rev. {\bf D35}, 2769(1987); J. Flynn and L. Randall,
Nucl. Phys. {\bf B326}, 31(1989); Erratum-ibid, {\bf B334}, 580(1990).

\bibitem{isidori} G. Buchalla, G. D'Ambrosio, and G. Isidori,
Nucl. Phys. {\bf B672}, 387(2003); G. Isidori, C. Smith and R. Underdorfer,
Eur. Phys. J. {\bf C36}, 57(2004).


\bibitem{na48} J. Batley et al., NA48 Collaboration, Phys. Let. {\bf 576},
43(2003); ibid, {\bf B599}, 197(2004).

\bibitem{buras2} A. Buras, F. Schwab and S. Uhlig, arXiv: hep-ph/0405132.

\bibitem{buras3} G. Buchalla, and A. Buras, Nucl. Phys. {\bf B398}, 285(1993);
225(1993); {\bf B548}, 309(1999);
M. Misiak and J. Urban, Phys. Lett. {\bf B451}, 161(1999).

\bibitem{marciano} W.J. Marciano and Z. Parsa, Phys. rev. {\bf D53}, 1(1996).

\bibitem{e787} S. Adler et al., E787 Collaboration, Phys. Rev. Lett.
{\bf 88}, 041803(2002).

\bibitem{e949} A.V. Artamonov et al., E949 Collaboration, arXiv: hep-ex/0403036.


\bibitem{beyondsm}
G. Colangelo and G. Isidori, JHEP {\bf 9809}, 009(1998);
A. Buras, A. Romanino and L. Silvestrini, Nucl. Phys. {\bf B520}, 3(1998);
A. Buras et al., Nucl. Phys. {\bf B566}, 3(2000);
S. Baek, J. Jang, P. Ko and J. Park, Nucl. Phys. {\bf B609}, 442(2001);
Y. Nir and G. Raz, Phys. rev. {\bf D66}, 035007(2002);
C.H. Chen, J. Phys. {\bf G28}, L33(2002);
W. F. Chang and J. Ng, JHEP {\bf 0212}, 077(2002);
G. D'Ambrosio, F. Giudice, G. Isidori and A. Strumia, Nucl. Phys. {\bf B645},
155(2002); G. Burdman, Phys. Rev. {\bf D66}, 076003(2002);
D. Hawkins and D. Silverman, Phys. Rev. {\bf D66}, 016008(2002);
T. Yanir, JHEP {\bf 0206}, 044(2002);
A. Buras, M. spranger and A. Weiler, Nucl. Phys. {\bf B660}, 225(2003);
A. Buras, R. Fleischer, S. Recksiegel and F. Schawb,
arXiv: hep-ph/0402112.



\bibitem{mfv} A. Buras et al., Nucl. Phys. {\bf B592}, 55(2001).


\bibitem{gsusy} A. Buras, T. Ewerth, S. Jager and J. Rosiekm arXiv:
hep-ph/0408142.

\bibitem{rparity}
N. G. Deshpande, D. K. Ghosh, and X.-G. He, Phys. Rev. {\bf D70},
093003(2004)[arXiv: hep-ph/0407021].

\bibitem{rparity1}
A. Deandrea, J. Welzel and M. Oertel, JHEP {\bf 0410},
038(2004)[arXiv: hep-ph/0407216].

\bibitem{hv} X.-G. He and G. Valencia, Phys. Rev. {\bf D70}, 053003(2004)[arXiv:
hep-ph/0404229].

\bibitem{little} S. Choudhury, N. Gaur, G. Joshi and B. McKellar,
arXiv: hep-ph/0408125.

\bibitem{isosinglet}
J. Aguilar-Saavedra, Phys. Rev. {\bf D67}, 035003(2003).

\bibitem{lrmodel} Z. Gayi-Palffy, A. Pilaftsis and K. Schilcher,
Phys. Lett. {\bf B343}, 275(1995); Nucl. Phys. {\bf B513},
517(1998); X.-G. He, G. Valencia and Y.-L. Wang, Phys. Rev. {\bf
D70}, 11301(2004)[arXiv: hep-ph/0409346].

\bibitem{gross} Y. Grossman, G. Isidori and H. Murayama,
Phys. Lett. {\bf B588}, 74(2004).



\end{thebibliography}
